\documentclass[twocolumn]{aastex61}

\received{2017 April 24}
\revised{2017 October 3}
\accepted{2017 October 4}

\shorttitle{polarized lunar eclipse spectrum}
\shortauthors{Takahashi et al.}

\begin{document}

\title{Polarized Transmission Spectrum of Earth as Observed during a Lunar Eclipse}

\correspondingauthor{Jun Takahashi}
\email{takahashi@nhao.jp}

\author{Jun Takahashi}
\affil{Center for Astronomy, University of Hyogo, 407-2 Nishigaichi, Sayo, Hyogo 679-5313, Japan}

\author{Yoichi Itoh}
\affil{Center for Astronomy, University of Hyogo, 407-2 Nishigaichi, Sayo, Hyogo 679-5313, Japan}

\author{Kensuke Hosoya}
\affil{Center for Astronomy, University of Hyogo, 407-2 Nishigaichi, Sayo, Hyogo 679-5313, Japan}

\author{Padma A. Yanamandra-Fisher}
\affiliation{Space Science Institute, 4750 Walnut St, Suite 205, Boulder, CO 80301, USA}

\author{Takashi Hattori}
\affiliation{Subaru Telescope, National Astronomical Observatory of Japan, 650 North A'ohoku Place, Hilo, HI 96720, USA}



\begin{abstract}

Polarization during a lunar eclipse is a forgotten mystery.
\cite{coyn1970} reported the detection of significant polarization {during} the lunar eclipse on 1968 April 13.
Multiple scattering during the first transmission through the Earth's atmosphere was suggested as a possible cause of the polarization, but no conclusive determination was made.
{No} further investigations on polarization during a lunar eclipse {are known}.
We revisit this mystery with an interest in possible application to extrasolar planets;
if planetary transmitted light is indeed polarized, it may be possible to investigate an exoplanet atmosphere using ``transit polarimetry.''
Here we report results of the first spectropolarimetry for the Moon during a lunar eclipse on 2015 April 4.
We observed polarization degrees of 2--3 \%  at wavelengths of 500--600 nm;  in addition, an enhanced feature was detected at the O$_2$ A band near 760 nm.
The observed time variation and wavelength dependence are consistent with an explanation of polarization caused by double scattering during the first transmission through the Earth's atmosphere, accompanied by latitudinal atmospheric inhomogeneity.
Transit polarimetry for exoplanets may be useful to detect O$_2$ gas and to probe the latitudinal atmospheric inhomogeneity, and it is thus worthy of serious consideration.

\end{abstract}




\section{Introduction} \label{sec:intro}

\cite{coyn1970} reported the detection of a $\sim$2 \% polarization degree of the integrated lunar disk during the eclipse on 1968 April 13. 
According to them, polarization of the Moon during a lunar eclipse was reported at least twice before their detection.
However, polarization of the eclipsed Moon is puzzling.
A lunar eclipse takes place when the Sun, the Earth, and the Moon are nearly aligned. 
In general, the emergence of polarization requires some sort of anisotropy with respect to the orientation of electromagnetic oscillation.
Hence, such a straight optical path is hardly considered to produce polarization.

The eclipsed Moon is illuminated by sunlight transmitted through the Earth's atmosphere.
Scattered light in the atmosphere may contribute to the brightness of the eclipsed Moon \citep{garc2011a,garc2011b}, but single scattering with a scattering angle less than $\sim$2$^\circ$ cannot account for the $\sim$2 \% polarization degree reported by \cite{coyn1970}.
They suggested multiple scattering, namely, double scattering, as a possible cause of polarization, though they did not reach a conclusion. 
To our knowledge, no further investigations on this phenomenon have been reported.

In light of recent progress in the science of extrasolar planets (exoplanets), lunar eclipse observations have been given a new role.
The geometry of a lunar eclipse resembles that of an exoplanet transiting in front of its host star.
Thus, lunar eclipses provide ground-based observers with opportunities to observe the Earth as a transiting exoplanet and to obtain Earth's transmission spectra \citep{pall2009,vida2010, yan2015}.
These results are valuable to assess the scientific importance and feasibility of future transit spectroscopy for Earth-like exoplanets.
It should be noted that transit spectroscopy (or multiwavelength photometry) has already been conducted for giant exoplanets; 
some atmospheric species were successively detected \citep[e.g.,][]{tine2007,mccu2014}.

If transmission through a planetary atmosphere indeed polarizes incident stellar light, we may have ``transit polarimetry'' within our grasp as a new method to investigate the atmospheric properties of planets, because some information of the atmosphere must be imprinted into the polarization of the transmitted light.
In order to explore the potential of transit polarimetry, we need to confirm and understand polarization during a lunar eclipse.

The rest of this article reports the results of our first spectropolarimetry for a lunar eclipse.
Sections \ref{sec:obs} and \ref{sec:reduce} describe our observations and data reduction method, respectively.
The main results of our observations are provided  in Section \ref{sec:result}.
Some possible causes of the observed polarization are discussed in Section \ref{sec:discuss}.
We give conclusions for this article in Section \ref{sec:conclude}.
Supplementary information with respect to the data recuction and the reliability of our results is provided in Appendixes \ref{apx:othersky} amd \ref{apx:credibility}.
{A} detailed discussion {of} the first transmission through the Earth's atmosphere as one of the possible polarizing processes is given in Appendix \ref{apx:detail}.  
A brief discussion of the expected fractional polarization during a transit of an Earth-like exoplanet is presented in Appendix \ref{apx:estimate}.


\section{Observations} \label{sec:obs}

\begin{figure}[htbp]
\begin{center}
\includegraphics[width=\linewidth]{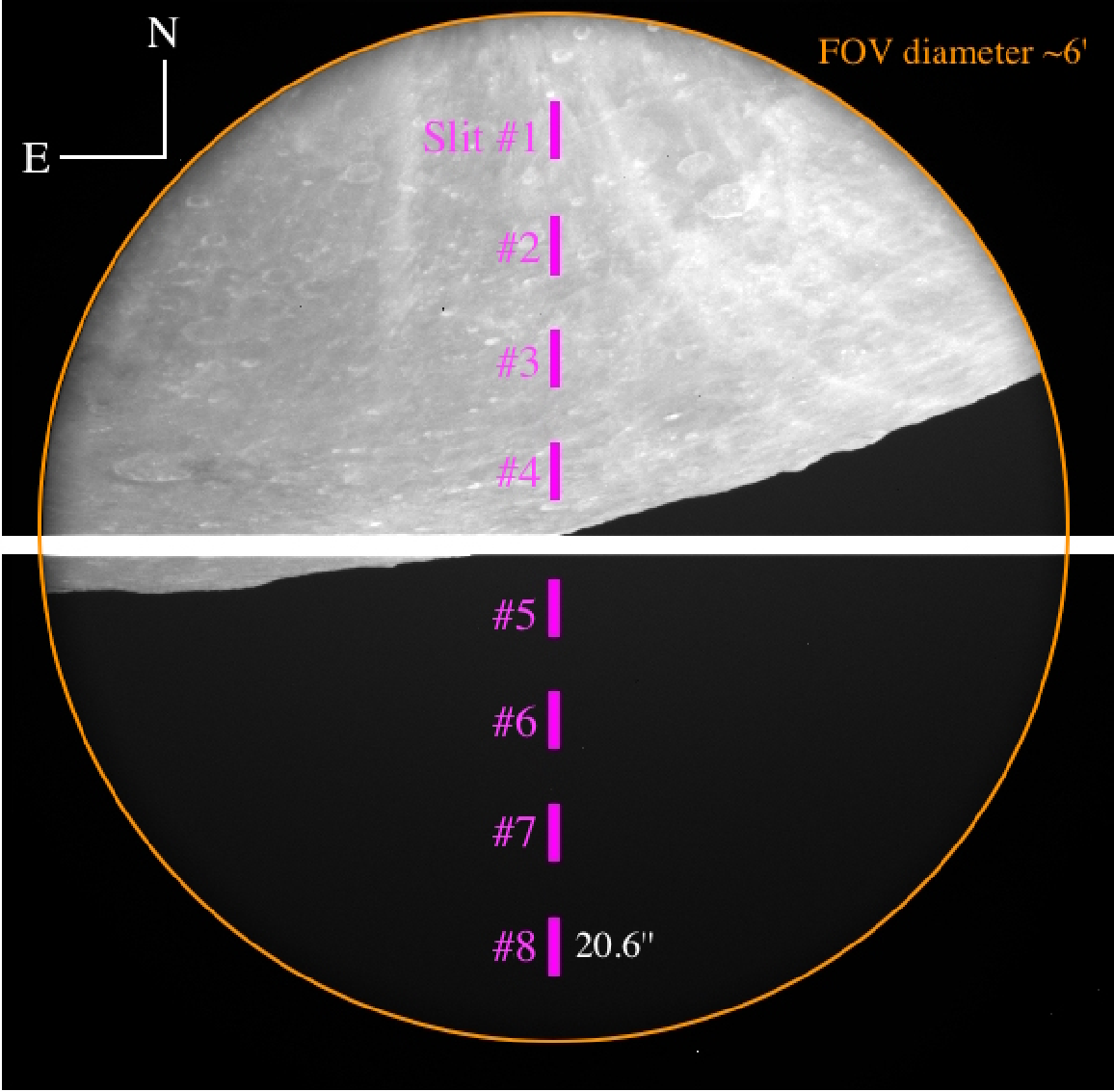}
\caption{Observed region on the Moon and placement of the 2.0$''$-wide slits.
This image was taken through the imaging mode of FOCAS.
The diameter of the field of view is $\sim$6$'$.
The FOCAS detector consists of two chips; 
there is a $\sim$7$''$ gap between the chips.
The placement of the 0.4$''$ slits is almost the same as that of the 2.0$''$ slits, but shifted slightly to the north by $\sim$8.5$''$.}
\label{fig:slits}
\end{center}
\end{figure}

We conducted the first spectropolarimetry for the eclipsed Moon on 2015 April 4, with an anticipation that the wavelength dependence and time variation of polarization would facilitate identification of the cause.
The observations were made using the Faint Object Camera and Spectrograph  \citep[FOCAS;][]{kash2000} mounted on the 8.2 m Subaru Telescope, located at Maunakea, Hawaii.
The field of view was centered at Le Gentil A crater (lunar coordinates: 
52.4$^\circ$W, 74.6$^\circ$S) near the southern edge of the lunar disk.
The FOCAS is equipped with eight 20.6$''$-long slits, which are aligned in line with separations of 53.6$''$ (between the central two slits) {and} 44.3$''$ (between the others).
The slits were kept in the north--south orientation throughout the observations.
The northern four slits were placed on the Moon, whereas the other four were on the sky background (Figure  \ref{fig:slits}).
We selected  2.0$''$-wide slits for in-umbra observations, and 0.4$''$ for out-of-umbra observations, respectively.
The grating and filter setting was ``300B'' + ``SY47,'' which provided a wavelength ($\lambda$) coverage of 500--800 nm and  a wavelength resolution ($\lambda/\Delta \lambda$) of $\sim$200 for the 2.0$''$ slits or $\sim$1000 for  the 0.4$''$ slits.

Light obtained through a slit is split by a Wollaston prism into ordinary and extraordinary spectra.
{A single dataset consists} of four images obtained thorough a half-wave plate with its position angles of 0$^\circ$, 45$^\circ$, 22.5$^\circ$, and 67.5$^\circ$.

The partial eclipse began at 2015 April 4, 10:15 (UT), and the total eclipse occurred from 11:54 to 12:06 (Table \ref{tbl:event}). 
The target region was in the umbra from $\sim$10:40 until $\sim$13:20. 
We continued observing the same region from $\sim$10:30 until $\sim$13:30 (Table \ref{tbl:obslog}).
However, it became cloudy after the greatest eclipse, and thus we could not acquire effective data after the totality until the target's egress from the umbra.

\begin{table}[htbp]
\caption{Timetable of the events associated with the lunar eclipse on 2015 April 4.}
\begin{center}
\begin{tabular}{rl}
\hline
\hline
Time (UT) &  Event \\
\hline
10:15 &  Beginning of partial eclipse \\ 
$\sim$10:40 &  Ingress of target region into umbra \\
11:54 &  Beginning of total eclipse\\
12:00 &  Greatest eclipse \\
12:06 &  End of total eclipse \\
$\sim$13:20 &  Egress of target region out of umbra \\
13:45 &  End of partial eclipse \\
\hline
\end{tabular}
\end{center}
\label{tbl:event}
\tablecomments{The times with respect {to the} partial/total eclipse are the predictions by the Ephemeris Computation Office of the National Astronomical Observatory of Japan.}
\end{table}%

\begin{table*}[htbp]
\caption{Log of the observations.}
\begin{center}
\begin{tabular}{ccclcl}
\hline
\hline
Time &  Out-of-/In-umbra & Slit Width & Exposure Times & Num. Sets\tablenotemark{a} & Remarks \\
 (UT) &   &  &  (sec) &   &  \\

\hline
10:27--10:35 &  Out-of & 0.4$''$ & 0.3  & 5 &  \\ 
10:35--10:42 &  Transition & 0.4$''$ & 1.0, 2.0, 5.0  & 4 &  \\ 
10:43--11:11 &  In & 2.0$''$ & 1.0, 2.0, 3.0  & 10 &  \\ 
11:12--{12:10} &  In & 2.0$''$ & 4.0, 5.0, 8.0, 10.0   & {25} &  \\ 
{12:11}--13:22 &  In & 2.0$''$ & 1.0--30.0  & {28} & Cloudy \\ 
13:25--13:28 &  Out-of & 0.4$''$ & 2.0, 3.0  & 2 & Cloudy\tablenotemark{b} \\ 
\hline
\end{tabular}
\end{center}
\label{tbl:obslog}
\tablenotetext{a}{A single set consists of four exposures taken with four different position angles of the half-wave plate.}
\tablenotetext{b}{Despite clouds, the brightness of the out-of-umbra Moon allowed us to obtain sufficient signals.
Thus, the results are included in Figures \ref{fig:PDtime} and \ref{fig:PAtime}.}
\end{table*}%

\section{Data Reduction} \label{sec:reduce}

Bias counts in the raw data were subtracted using the images obtained without an exposure.
Overscan regions on an image were trimmed using an IRAF \cite[Image Reduction and Analysis Facility;][]{tody1986,tody1993} task called ovsub\_hpk in FOCASRED,\footnote{\url{https://subarutelescope.org/Observing/Instruments/FOCAS/Detail/UsersGuide/DataReduction/focasred.html}} which is a software package prepared for FOCAS data.
Flat-fielding was conducted with the use of dome flat images, which were obtained through a half-wave plate with its position angles of 0$^\circ$, 45$^\circ$, 22.5$^\circ$, and 67.5$^\circ$.
The images with four position angles were combined to synthesize an unpolarized flat image.

Wavelength calibration was performed by referring to known emission lines on Th-Ar arc spectra.
A two-dimensional (function of position and wavelength) spectral image obtained thorough a slit was averaged into a one-dimensional (wavelength) intensity spectrum.
Because the red end of spectra obtained through slits \#1 and \#8 were found to suffer from vignetting, data from these slits were excluded.

\begin{figure*}[htbp]
\begin{center}
    \begin{tabular}{cc}
     \resizebox{0.5\hsize}{!}{\includegraphics{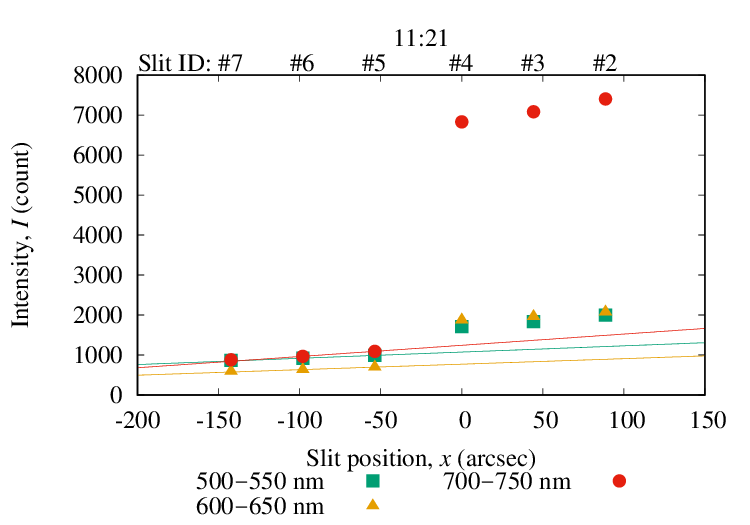}}  &
     \resizebox{0.5\hsize}{!}{\includegraphics{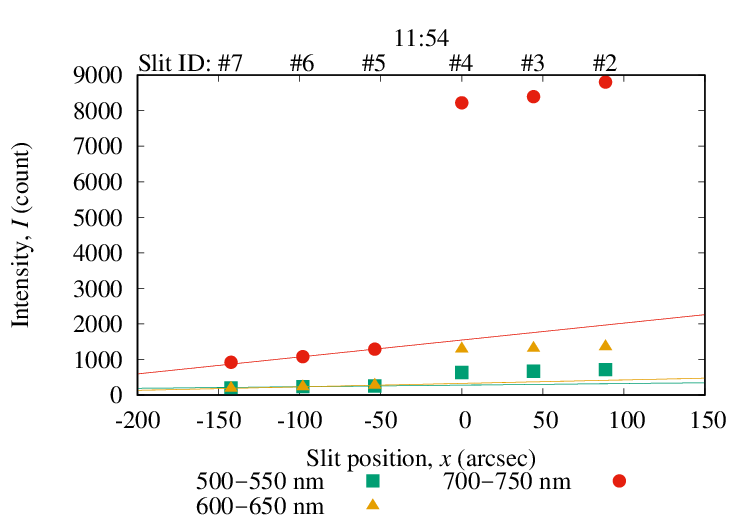}} 
     \end{tabular}
\caption{Extrapolation of sky background. 
Data for single exposures at UT 11:21 (left) and 11:54 (right) are shown.
Slits \#2--4 are on the Moon, whereas slits \#5--7 are on the sky.
Although the fitting and subtraction were {performed} spectrally (at each wavelength element), the intensity values (the points) and fit lines are shown in terms of averages in the three wavelength ranges, for clarity.}
\label{fig:skyfit}
\end{center}
\end{figure*}

The intensity of sky background (which includes the light scattered during transmission through the telluric atmosphere on the path from the Moon to the telescope) was subtracted as described below.
The sky intensity is greater when a slit position is nearer to the Moon (Figure  \ref{fig:skyfit}).
The dependence of the intensity {on} the slit position is roughly linear.
Thus, we linearly extrapolated the intensity spectra of the sky (slits \#5 and \#6) to the slit positions on the Moon (slits \#4, \#3, and \#2).
The estimated sky intensities on the lunar positions were subtracted from raw spectra of the Moon.
Note that the subtraction was conducted spectrally (at each wavelength element) and separately for both the ordinary and extraordinary spectra.

\begin{figure*}[htbp]
\begin{center}
    \begin{tabular}{cc}
    \resizebox{0.5\hsize}{!}{\includegraphics[width=\linewidth]{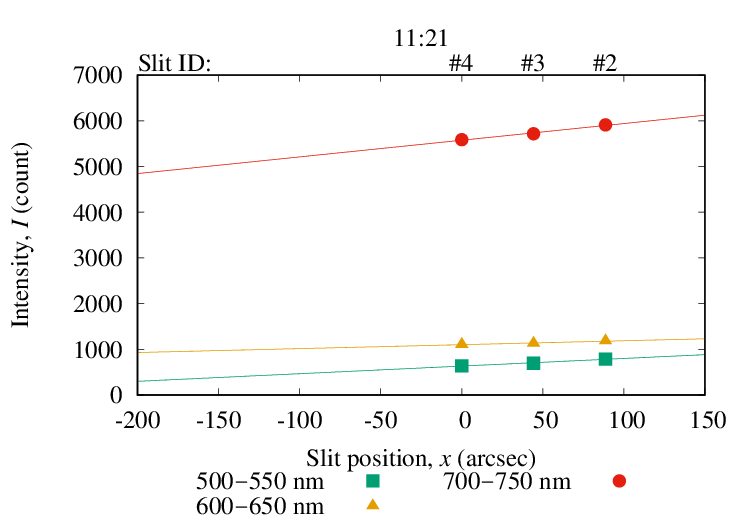}} &
    \resizebox{0.5\hsize}{!}{\includegraphics[width=\linewidth]{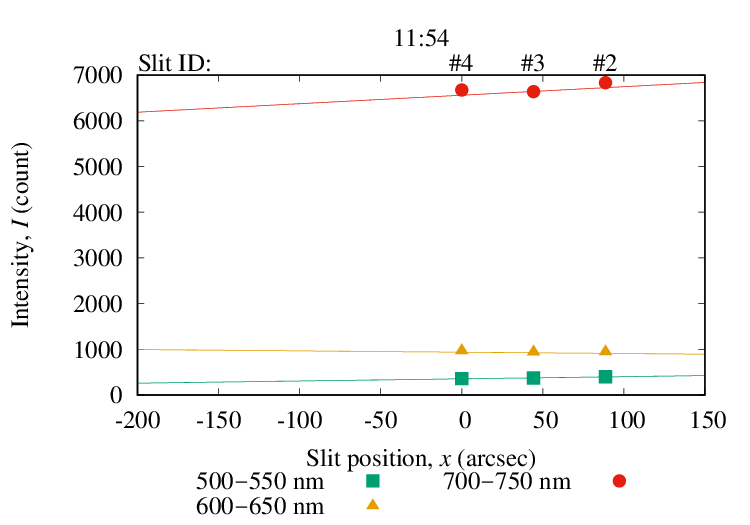}}
     \end{tabular}
\caption{Residual fitting.
{The points} are the intensities after the first sky subtractions.
The lines are those for the residual fitting.
Although the intensities after the first sky subtraction for UT 11:54 (right) are fairly flat, those for UT 11:21 (left) present clear monotonic increases with increasing $x$. }
\label{fig:residual}
\end{center}
\end{figure*}

The sky-subtracted intensity spectra on the Moon were, however, found to still have a systematic dependence of intensity on the slit position (Figure \ref{fig:residual}).
This is probably due to the fact that the positional dependence of sky intensity is not exactly linear, and thus linear subtraction leaves a certain {number} of residual  {counts}.
Therefore, we again performed a fitting of a linear function $\alpha x + \beta$ to the sky-subtracted Moon spectra (on slits \#4, \#3, and \#2), as shown in Figure \ref{fig:residual}, and then evaluated the residual counts by $\alpha x$. 
Note that $\alpha$ is the slope of the fitted line and a function of wavelength.
Subtraction of the residual counts reduced the systematic dependence of derived polarization degrees on slit position.

Although we made a conscientious effort to treat the sky subtraction, we will keep in mind  the imperfection of  our method  and try to evaluate its impact later in this section.
{In addition, we try other sky subtraction techniques in Appendix \ref{apx:othersky} and check the reliability of our results from different perspectives in Appendix \ref{apx:credibility}.}

{Stokes parameters (ratios) $q_0=Q/I$ and $u_0=U/I$} were derived from a set of intensity spectra observed with position angles of the half-wave plate of 0$^\circ$, 45$^\circ$, 22.5$^\circ$, and 67.5$^\circ$ as follows:
\begin{eqnarray}
a_q & = & 
\sqrt{\frac{I_\textrm{e}(0^\circ)}{I_\textrm{o}(0^\circ)} 
 \frac{I_\textrm{o}(45^\circ)}{I_\textrm{e}(45^\circ)}}, \label{eq:aq}\\
a_u & = & 
\sqrt{\frac{I_\textrm{e}(22.5^\circ)}{I_\textrm{o}(22.5^\circ)} 
\frac{I_\textrm{o}(67.5^\circ)}{I_\textrm{e}(67.5^\circ)}}, \label{eq:au}\\
q_0 & = & \frac{1-a_q}{1+a_q},\\
u_0 & = & \frac{1-a_u}{1+a_u},
\end{eqnarray}
where $I(\phi)$ represents intensity observed through a half-wave plate with its position angle of $\phi$.  
{The subscripts} ``o'' and ``e'' stand for ordinary and extraordinary rays, respectively, which are split by a Wollaston prism.
Note that calculations in equations (\ref{eq:aq}) and (\ref{eq:au}) cancel out the telluric transmittance during a single exposure for $I_o(\phi)$ and $I_e(\phi)$; hence, in principle, the time variation of telluric transmittance between separate exposures does not alter $q_0$ or $u_0$.
{The instrumental} polarizations ($q_\textrm{inst}$, $u_\textrm{inst}$) were measured by observing unpolarized standard stars (HD 94851 and HD 154892) and subtracted from $q_0$ and $u_0$ by
$q_1  =  q_0 - q_\textrm{inst}$ and $u_1  =  u_0 - u_\textrm{inst}$, respectively.
Note that the instrumental polarization depends on the slit position.
Hence, we obtained $q_\textrm{inst}$ and $u_\textrm{inst}$ for each slit.
For the purpose of reducing random errors, the median values $(\bar{q}_1, \bar{u}_1)$ were taken within several sets of successively observed $(q_1, u_1)$ data (two or three sets for the out-of-umbra spectra and five sets for the in-umbra spectra), which resulted in 11 pairs of $(\bar{q}_1, \bar{u}_1)$ median spectra in total, out of 46 $(q_1, u_1)$ datasets.
The polarization degree ($P$) and position angle of polarization ($\Theta_0$) were derived by
\begin{eqnarray}
P & = & \sqrt{\bar{q}_1^2 + \bar{u}_1^2},\\
\mathrm{tan}\; 2\Theta_0 & = & \bar{u}_1/\bar{q}_1, \\
\mathrm{sgn}(\mathrm{cos}\; 2\Theta_0 ) & = & \mathrm{sgn}\; \bar{q}_1, 
\end{eqnarray}
where sgn represents ``sign of.''
The position angle of polarization was corrected by $\Theta = \Theta_0 - \Delta \Theta$, 
where $\Delta \Theta$ is the difference {in} position angle origins between the instrumental and equatorial coordinates, which was measured using a strongly polarized standard star (HD 43384). 
Although the $P$ and $\Theta$ spectra for all of slits \#2, \#3, and \#4 were derived, we adopted the values from slit \#4 as the final results.
This is because slit \#4 is the nearest to the sky (Figure \ref{fig:slits}), and thus the estimated (extrapolated) sky intensity for slit \#4 should be more accurate than those for slits \#3 and \#2.
The values from the other two slits were used to estimate the errors (see below).

Random errors of $q_1$ and $u_1$ ($\sigma_q^\mathrm{rand}$ and  $\sigma_u^\mathrm{rand}$) were calculated as rms within several sets of successively observed $q_1 - \bar{q}_1$ and $u_1 - \bar{u}_1$.  
The typical level of systematic errors depending on the slit position ($\sigma_q^\mathrm{slit}$ and  $\sigma_u^\mathrm{slit}$) was evaluated as rms of $\bar{q}_1 (x_n) -  \bar{q}_1(0) $ and $\bar{u}_1 (x_n) -  \bar{u}_1(0) $, where $x_n$ is the position of slit \#$n$ with respect to the position of slit \#4 ($\bar{q}_1(0)$ and $\bar{u}_1(0)$ are those derived from slit \#4).
The errors $\sigma_q^\mathrm{slit}$ and  $\sigma_u^\mathrm{slit}$ include those by the imperfect sky subtraction and inhomogeneity in depolarization factors on the lunar surface because these effects cause differences in $\bar{q}_1$ and $\bar{u}_1$ for different  slits.
The error propagation to $P$ from $q$ and $u$ was determined by
\begin{eqnarray}
\sigma_P & = & \frac{\sqrt{q^2\sigma_q^2+u^2\sigma_u^2}}{P},
\end{eqnarray}
where
\begin{eqnarray}
\sigma_q & = & \sqrt{(\sigma_q^\mathrm{rand})^2+(\sigma_q^\mathrm{slit})^2},\\
\sigma_u & = & \sqrt{(\sigma_u^\mathrm{rand})^2+(\sigma_u^\mathrm{slit})^2}.
\end{eqnarray}
In addition, we considered the imperfection in the correction of instrumental polarization ($\delta_P$), which was measured as the difference between the observed $P$ of the strongly polarized standard star (after correction of instrumental polarization) and  $P$ of the star in literature \citep{hsu1982}.
After all, the total errors in $P$ and $\Theta$ were evaluated by
\begin{eqnarray}
\varepsilon_P & = & \sqrt{\sigma_P^2+\delta_P^2}, \label{eq:errP}\\
\varepsilon_\Theta & = & 28.65^\circ \frac{\varepsilon_P}{P}. \label{eq:errTheta}
\end{eqnarray}

\section{Results} \label{sec:result}

\begin{figure*}
\centering
    \begin{tabular}{cc}
     \multicolumn{1}{l}{(a)}  &      \multicolumn{1}{l}{(b)}\\
     \resizebox{0.5\hsize}{!}{\includegraphics{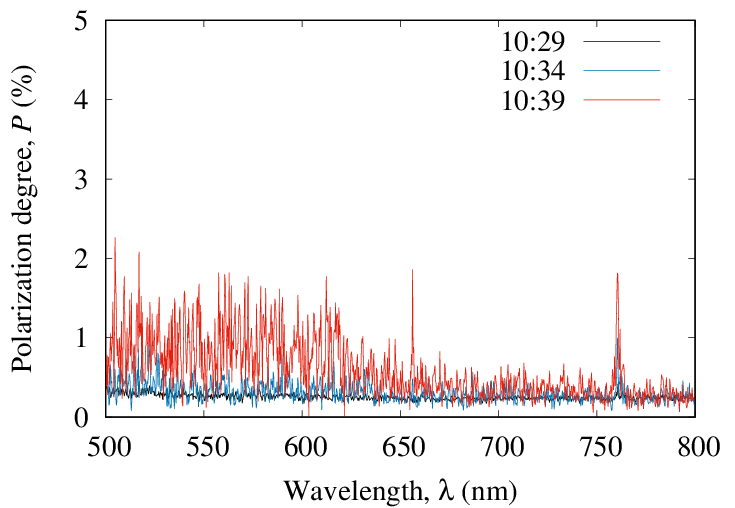}}  &
      \resizebox{0.5\hsize}{!}{\includegraphics{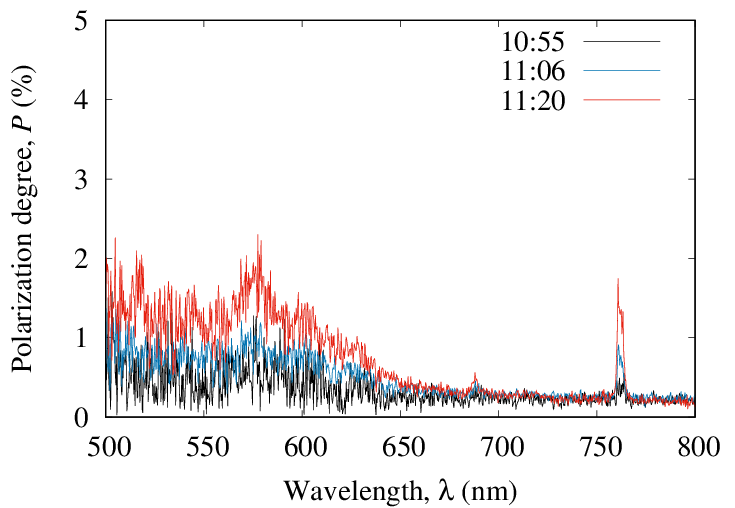}}  \\ \\
     \multicolumn{1}{l}{(c)} & \multicolumn{1}{l}{(d)} \\
      \resizebox{0.5\hsize}{!}{\includegraphics{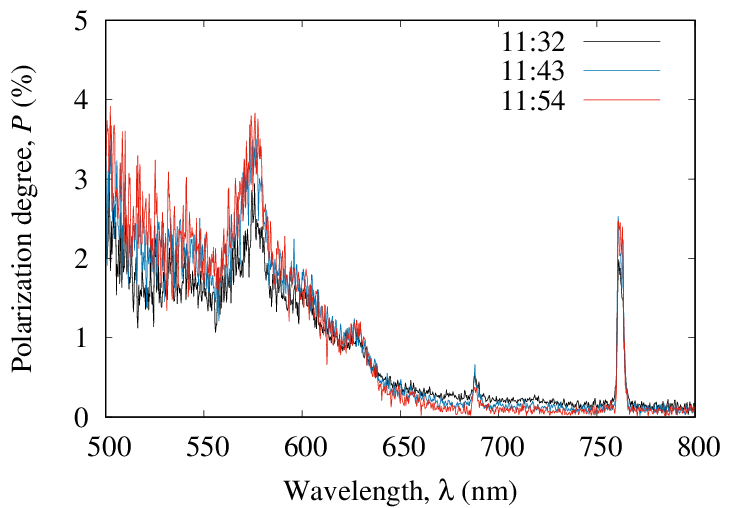}}  &
      \resizebox{0.5\hsize}{!}{\includegraphics{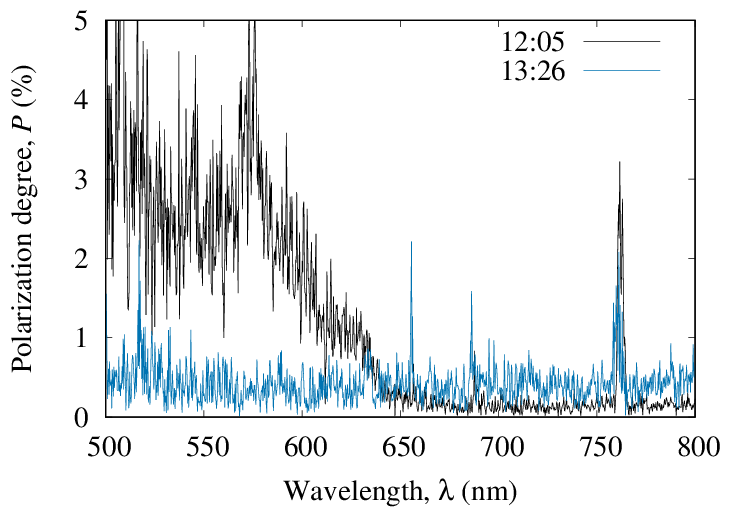}}
    \end{tabular}
    \caption{All of the polarization degree spectra obtained during the lunar eclipse on  2015 April 4. The average UT of observations for each spectrum is shown in the panels.}
    \label{fig:allspec}
\end{figure*}

\begin{figure}[htbp]
\begin{center}
\includegraphics[width=\linewidth]{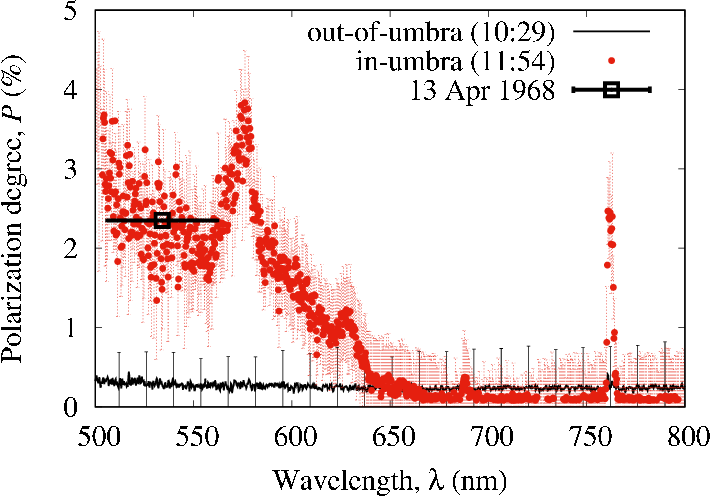}
\caption{{Out-of- and in-umbra polarization} degree spectra obtained during the lunar eclipse on  2015 April 4.
The black line represents a polarization degree spectrum when the target region on the Moon was outside of the Earth's umbra (the central time of the observations is 10:29 UT).
The red filled circles represent that when the target was inside of the umbra (the central time is 11:54, which is near the time of the greatest eclipse, 12:00).
Note that the error bars are exhibited only for every 10 points for the out-of-umbra data (10:29) and every two for the in-umbra data (11:54), for clarity.
The open square shows the result of polarimetry for a lunar eclipse on 1968 April 13 \citep{coyn1970}. 
}
\label{fig:PDspec}
\end{center}
\end{figure}

{All of the} deduced polarization degree spectra are displayed in Figure {\ref{fig:allspec}}.
In Figure \ref{fig:PDspec}, we also show a typical spectrum observed when the target region was outside of  the Earth's umbra (hereafter the ``out-of-umbra'' spectrum) and one obtained when the target region was inside of the umbra (the ``in-umbra'' spectrum), together with the errors defined by equation (\ref{eq:errP}).
The {out-of-umbra spectrum} is almost unpolarized over the whole wavelength range, whereas the {in-umbra spectrum} is significantly polarized for wavelengths shorter than 600 nm.
The in-umbra polarization degrees at wavelengths 500--600 nm are {approximately} 2--3 \%, which agrees with the report by \cite{coyn1970} (2.4 \% in a broad band of  534 $\pm$ 28 nm).
On the in-umbra spectrum, a strongly enhanced feature presents near 760 nm, which corresponds to oxygen (O$_2$) absorption wavelengths called the A band.  
Furthermore, another broader enhanced feature is noticeable at 560--580 nm, which corresponds to the absorption wavelengths of oxygen collision complexes \citep[O$_2\cdot$O$_2$, or O$_2$ dimer;][]{pall2009}.
Although buried in the error bars, two more features appear near 630 and 690 nm, which may also correspond to O$_2$ absorption wavelengths.

\begin{figure}[htbp]
\begin{center}
\includegraphics[width=\linewidth]{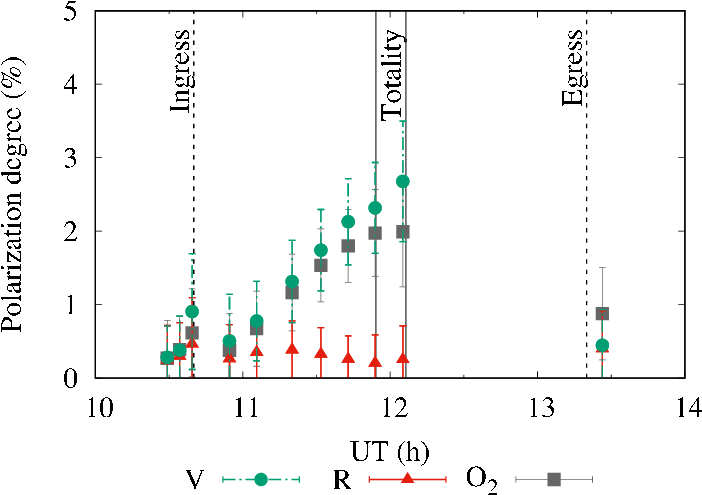}
\caption{Time variation of polarization degree.
{The green} circles, red triangles, and gray squares represent polarization degrees averaged within the V-band (550.5 $\pm$ 41.4 nm), R-band (658.8 $\pm$ 78.4 nm), and O$_2$ A-band (762.5 $\pm$ 1.5 nm) wavelengths, respectively, where the signal counts are used as weights in the averaging.
The vertical dashed lines show the approximate times of the target's ingress to and egress from the Earth's umbra.
The vertical solid lines indicate the beginning and end times of the totality.
Owing to clouds, we were not able to obtain effective data after the end of totality until the target egress (the errors in the V band are larger than 1.0 \%).
}
\label{fig:PDtime}
\end{center}
\end{figure}

The time variation of the polarization degree is shown in Figure \ref{fig:PDtime} for all of the derived  $P$ spectra.
The polarization degree in the V band (550.5 $\pm$ 41.4 nm) increased from the ingress time toward the mid-eclipse, whilst that in the R band (658.8 $\pm$ 78.4 nm) remained unpolarized. 
The time variation in the V band  seems consistent with the observation by \cite{coyn1970}, who reported that the polarization degree of the whole lunar disk decreased from 2.35 \% during the totality to $\sim$0.5 \% at the time of fourth contact (when the whole Moon exited from the umbra). 
They could not observe the ingress, owing to clouds.
The polarization degree in the O$_2$ A band (averaged within 762.5 $\pm$ 1.5 nm) exhibited similar variation to that in the V band.

\begin{figure}[htbp]
\begin{center}
\includegraphics[width=\linewidth]{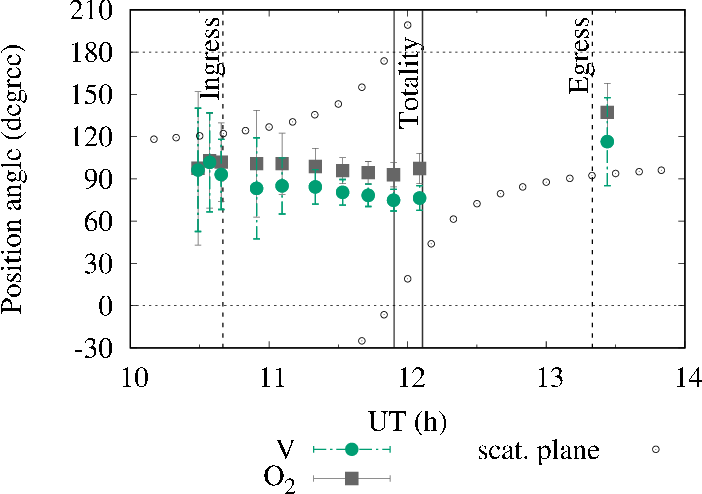}
\caption{Time variation of the polarization position angle.
{The position} angles are with respect to the equatorial north--south orientation and in a counterclockwise manner.
The same legends as in Figure \ref{fig:PDtime} are used.
The data for the R band are not shown because a significant polarization degree {was} not detected.
{The open} circles represent the approximate position angles of an on-sky line between the target position and the umbra center, which are identical to the position angles of the mean scattering plane defined by the Sun's center, the Earth's center, and the target position on the Moon.}
\label{fig:PAtime}
\end{center}
\end{figure}

Figure \ref{fig:PAtime} presents the time variation of the polarization position angle for all of the derived $\Theta$ spectra.
The position angle of the V-band polarization was virtually constant in $\sim90^\circ$, or the equatorial east--west direction.  
As shown in the figure, it appears to be independent of the position angle of the mean scattering plane defined by the Sun's center, Earth's center, and the target.
As is the case with the polarization degree, the position angle of polarization in the O$_2$ A band was similar to that in the V band.

\section{Discussion} \label{sec:discuss}
What caused the observed in-umbra polarization?
In a large sense, there are three different processes {from} which the observed in-umbra polarization may have arisen: (a) transmission through the Earth's atmosphere on the path from the Sun to the Moon, (b) reflection at the lunar surface, and (c) transmission through the Earth's atmosphere on the path from the Moon to the telescope.
Below, we discuss each process separately.
Sunlight transmits through the Earth's atmosphere twice before it is observed on the ground.
In order to distinguish the two different paths of transmission, we often refer to  (a) as the first transmission and  (c) as the second transmission.

\subsection{First Transmission through the Earth's Atmosphere}\label{sec:tans1}

In Appendix \ref{apx:detail}, we discuss in detail the first transmission as a cause of polarization.
Below, the essence of our discussion is summarized.

In analogy with Jupiter's limb polarization, we examine whether the Earth's limb can be polarized by double scattering during the first transmission.
We have constructed a simple model to evaluate the {doubly} scattered flux in Appendix \ref{apx:double}.
It shows that the doubly scattered flux emerges most effectively from a tangential altitude where the horizontal optical thickness is intermediate, or near unity (see Figure \ref{fig:polflux} in Appendix \ref{apx:detail}; we refer to this altitude as the boundary altitude), which is $\sim$15 km for a wavelength of 550 nm.
This can be interpreted as follows: 
within a horizontally thin layer, scattering becomes more effective with increasing atmospheric {density} (decreasing the altitudes).
However, when the layer becomes horizontally thick, the scattered flux going through the atmosphere toward the Moon decreases with increasing the {density} {owing} to increasing extinction. 

We examine an imbalance of fluxes emerging from the boundary altitude between beams that are scattered twice in a horizontal plane and those in a vertical plane (see Figure \ref{fig:AtmosScatter} in Appendix \ref{apx:detail} for a schematic illustration).
The horizontal optical thickness at a certain tangential altitude is $\sim$75 times larger than the vertical optical thickness above the same altitude \citep{fort2005}.
Hence, the horizontally scattered flux should be $\sim$75 times greater than the vertically scattered flux from the layer above the boundary altitude.
Our model shows that the vertically scattered flux from the layer below the boundary altitude is comparable to that from the layer above.
Consequently, the horizontally scattered flux is expected to overwhelm the vertically scattered flux (as the sum of those from both the upper and lower layers) by a factor of $\sim$30.
This results in polarization in the vertical orientation because scattered light is usually polarized perpendicularly to the scattering plane that includes the incident and scattered rays.

All of the overall wavelength dependence of the observed in-umbra polarization degree spectra (Figure \ref{fig:PDspec}), presence of the enhanced molecular features on the spectra (Figure \ref{fig:PDspec}), and their time variation (Figure \ref{fig:PDtime}) can be explained as the result of the combination of the polarized doubly scattered component and the unpolarized straightforward component (see Appendix \ref{apx:consist}).

During a lunar eclipse, the Earth is viewed as a ring of the bright limb from the Moon, as photographed by Surveyor III \citep{shoe1968} and  KAGUYA (SELENE)\footnote{\url{http://global.jaxa.jp/press/2009/02/20090218_kaguya_e.html}}.
We expect the Earth's limb to have a radial pattern of polarization position angles, because each element is estimated to be polarized vertically with respect to the limb.

Some sort of latitudinal inhomogeneity in the polarized flux is required in order for the integrated Earth's limb to be polarized.
Because the observed polarization position angles were nearly in the east--west orientation (Figure \ref{fig:PAtime}), we expect an excess of polarized flux at the equatorial regions compared to that at the polar regions. 
Although we are yet to identify the cause of the inhomogeneity, we point to a few possible causes: a greater scale height of the atmosphere above the equatorial region (according to our model, the polarized flux is proportional to the scale height), and/or a greater cloud coverage above the polar regions at the time of our observation (Appendix \ref{apx:double}).

\subsection{Reflection at the Lunar Surface}\label{sec:moon}

We examine the reflection at the lunar surface as another possible cause of the observed in-umbra polarization.
Light reflected from some of the airless solid bodies in the solar system is known to exhibit a polarization surge at very small phase angles less than $\sim$2$^\circ$ \citep[e.g.,][]{lyot1929,rose2005a,rose2005b,rose2009}.
The polarization degree at the surge is typically 0.3--0.5 \% \citep{rose2009}.
This effect, called the polarization opposition effect, may account for the observed polarization because the phase angle of the Moon ([the Sun's center]--[the Moon's center]--[observer] angle) was less than 1$^\circ$ during our observations.
According to a review by \cite{hapk2012}, the polarization opposition effect for the Moon has never been reported, though a laboratory measurement of a lunar soil sample detected the effect with a polarization degree of $\sim$0.5 \%.   

The observed wavelength dependence of the polarization opposition effect for silicate-surface bodies is available only for asteroid 64 Angelina  \citep{rose2005a} and Jupiter's satellite Io \citep{rose2005b}.
The results of multiband polarimetry for these two objects hardly showed a wavelength dependence {of} the polarization degrees;
the variation in the polarization degrees between wavelengths 500 and 800 nm was within a factor of {two} \citep{rose2005a, rose2005b}.
In contrast, the in-umbra polarization degrees in the same wavelength range present a variation {by} a factor of 10 or more (Figure \ref{fig:PDspec}).
Thus, the polarization opposition effect is inconsistent with our observed results with {respect} to the wavelength dependence {of the} polarization degrees.

Moreover, {because} the Moon has virtually no atmosphere, by no means can lunar reflection explain the presence of the strongly enhanced feature of the O$_2$ gas at wavelengths $\sim$760 nm.
The polarization properties at the V band and the O$_2$ A band are very similar in terms of the degrees, position angles, and time variations (Figures \ref{fig:PDspec}, \ref{fig:PDtime}, and \ref{fig:PAtime});
this suggests that the mechanisms causing the polarization in the two bands are common.
Therefore, the reflection at the Moon seems to have difficulty explaining polarization at the short-wavelength range (500--600 nm), as well as at the O$_2$ A band.

Considering the above-mentioned inconsistencies, we exclude reflection at the lunar surface as the cause of the in-umbra polarization.

\subsection{Second Transmission through the Earth's Atmosphere}\label{sec:trans2}
Finally, we discuss the second transmission through the Earth's atmosphere.
Double scattering does occur during the second transmission, as well as the first one, though a flux imbalance between polarization states should become less significant as the incidence geometry transitions from grazing to vertical.
However, the crucial difference between the two processes is the way of handling a scattering component.
As is the case of usual astronomical polarimetry through the ``second''-type transmission, a component of scattering within the telluric atmosphere is removed by the process of sky background subtraction (together with other background sources); 
we did conduct a sky subtraction procedure as described in Section \ref{sec:reduce}.
In contrast, a component of scattering during the first transmission is indistinguishably combined with the directly transmitted light; we count it as part of the true signals, not a contamination.

\subsection{Summary of the Discussion}

We have discussed the three possible causes of the observed in-umbra polarization: (a) the first transmission through the Earth's atmosphere on the path from the Sun to the Moon, (b) reflection at the lunar surface, and (c) the second transmission through the Earth's atmosphere on the path from the Moon to the telescope.
We have found (a) to be the most convincing.
The observed polarization can be explained by a horizontal/vertical imbalance in double scattering during the first transmission, accompanied by some sort of latitudinal inhomogeneity.
This explanation is consistent with the overall wavelength dependence, presence of molecular features, and the time variation of the observed polarization degree spectra.

\section{Concluding Remarks} \label{sec:conclude}
We conducted the first spectropolarimetry for the eclipsed Moon on  2015 April 4 and detected 2--3 \% polarization degrees at wavelengths 500--600 nm and at the O$_2$ A band near 760 nm.
The observed polarization is most probably caused by double scattering during the first Earth transmission.
This means that the transmitted light, as observed from the Moon or space, should be polarized by $\sim$6--9\% at 500--600 nm and the O$_2$ A band, given that lunar reflection depolarizes incident polarized light by a factor of $\sim$1/3 \citep{doll1957,bazz2013}.

Our findings imply that transit polarimetry may be used to investigate the atmospheres of {planets in the solar system and beyond, though detecting very small polarization during a transit of an Earth-like exoplanet is extremely challenging with current and near-future technologies (expected polarization degree of  $\sim 10^{-10}$; see Appendix \ref{apx:estimate})}.
For instance, the distinct feature at the O$_2$ A band can be used for detection of O$_2$ gas, which is a very important signature for the search {for} extraterrestrial life.
{The detection} of O$_2$ by the usual intensity spectroscopy is extremely challenging on the ground, owing to the contamination by the telluric O$_2$.
Polarimetry may be advantageous because the telluric spectral transmittance, including that at absorption wavelengths, is virtually isotropic with respect to orientation of electromagnetic oscillation, and thus it does not alter the polarization degree spectrum of astronomical objects. 
In addition to O$_2$ detection, transit polarimetry may become a unique method to probe the latitudinal inhomogeneity of {spatially unresolved planets} because it is required for the emergence of polarization during a transit; it is probably the type of information that cannot be accessed solely by intensity observations.
In order to evaluate the scientific importance and feasibility of transit polarimetry, further observational and theoretical studies are encouraged to fully understand polarization through planetary transmission.

\acknowledgments

This work is based on data collected at Subaru Telescope, which is operated by the National Astronomical Observatory of Japan.
We are grateful to Ji Hoon Kim for the support of our observations.
We express our gratitude to the anonymous referee for insightful comments and suggestions.
{This project is supported by} JSPS KAKENHI grant no.~15K21296.

%

\vspace{5mm}
\facilities{Subaru (FOCAS)}



\software{IRAF\citep{tody1986,tody1993},
FOCASRED}



\appendix

\section{Attempts of other sky subtraction techniques}\label{apx:othersky}

We explored the possibility of applying other sky subtraction techniques, which were previously established for lunar Earthshine observations.
\cite{taka2013} conducted Earthshine spectropolarimetry and developed a method in which a sky intensity spectrum is replaced by a spectrum for Moonshine (the bright side of the Moon) in order to gain a better signal-to-noise ratio (S/N).  
This method is made possible when the dominant source of sky intensity is Moonshine (i.e., the contribution from Earthshine and other light sources is negligible) and thus the sky spectrum is similar to that of Moonshine.
Unfortunately, this is not the case with lunar eclipse observations.  
The sky around an eclipsed Moon, especially during a partial eclipse, is contributed from light on both the in- and out-of-umbra portions of the Moon; 
their intensity spectra are quite different from each other, as is obvious from their colors viewed with the naked eye (in-umbra area is much redder).  
A sky spectrum around an eclipsed Moon is a mixture of different-shaped spectra and thus cannot be replaced by a single spectrum.
For this reason, we did not apply the method developed by \cite{taka2013}.

We also attempted a technique inspired by \cite{hamd2006} for lunar Earthshine intensity spectroscopy.
Just as we did, they linearly extrapolated the sky intensity toward the Moon to estimate the background intensity to be subtracted from the raw intensity of the Moon. 
They found that the slope for the linear extrapolation is proportional to the intensity of the sky.
Using this nature, they improve the S/N of the slope, especially at the wavelengths of strong absorption bands, by smoothing a spectrum of slope/intensity \citep[see Fig.\ 2 in][]{hamd2006}.
We checked a spectrum of slope/intensity for our data.
However, the slope/intensity spectrum was not flat and retained features at the absorption bands.
A simple mathematical examination showed that the slope is proportional to the intensity only when the sky intensity can be regarded as a sum of single-shaped spectra.
As described in the previous paragraph, this is the case for Earthshine observations, but not for lunar eclipse observations.

\section{Credibility of the Results}\label{apx:credibility}

Here, we check the credibility of the derived results.
Imperfect sky subtraction may lead to spurious results, because the sky intensities were comparable to the signals from the Moon, especially at the shortest wavelengths (Figure \ref{fig:skyfit}) and because the sky intensity spectra have features of O$_2$. 
As described in Section \ref{sec:reduce}, we accounted for the errors by the imperfect sky subtraction in our error evaluation.
The total errors are shown in Figure \ref{fig:PDspec}.
Considering the errors, the highest polarization degrees on the in-umbra Moon are justifiably significant.

\begin{figure}[htbp]
\begin{center}
\includegraphics[width=0.7\linewidth]{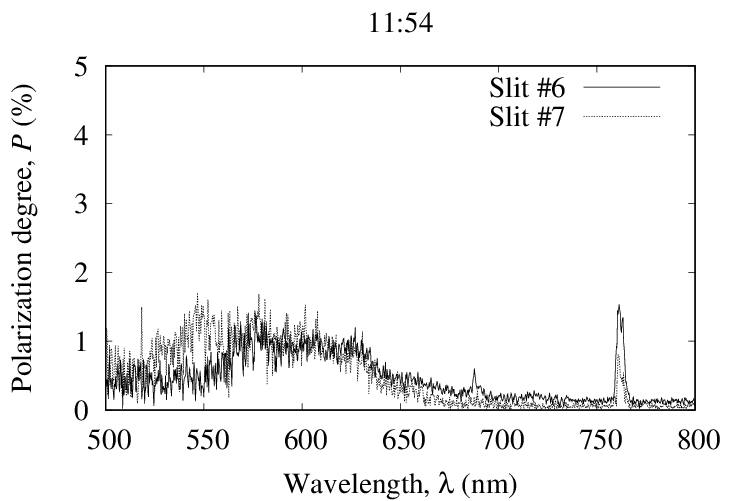}
\caption{Polarization degree spectra of the sky (11:54 UT).}
\label{fig:PDspecsky}
\end{center}
\end{figure}

\begin{figure}[htbp]
\begin{center}
\includegraphics[width=0.7\linewidth]{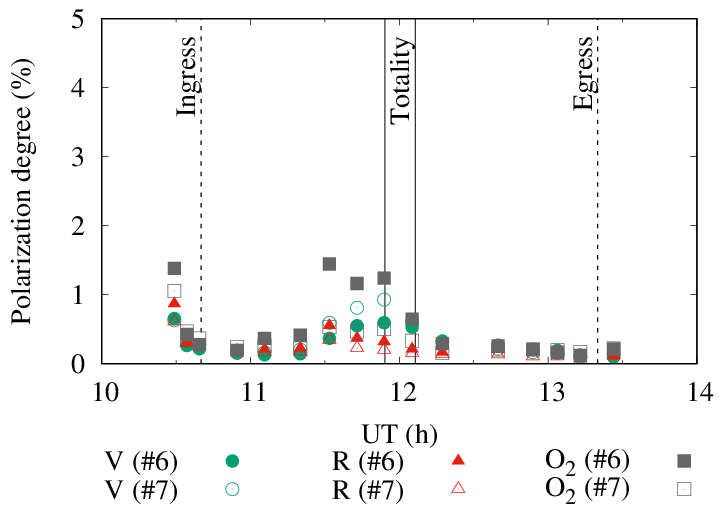}
\caption{Time-series polarization degrees of the sky.}
\label{fig:PDtimesky}
\end{center}
\end{figure}

\begin{figure}[htbp]
\begin{center}
\includegraphics[width=0.7\linewidth]{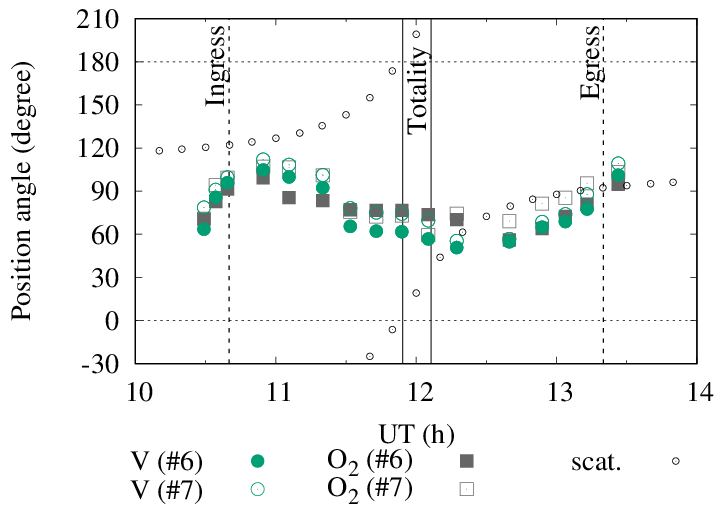}
\caption{Time-series polarization position angles of the sky.}
\label{fig:PAtimesky}
\end{center}
\end{figure}

In order to discuss the reliability of our results in a different manner, we measure the polarization of the sky itself at slits \#6 and \#7, where the instrumental polarization was derived by observing unpolarized standard stars through these slits. 
Its polarization degree spectra around the time of totality, the time-series polarization degrees, and the time-series position angles are shown in Figures \ref{fig:PDspecsky}, \ref{fig:PDtimesky}, and \ref{fig:PAtimesky}, respectively.
The polarization degree spectra of the sky have a similar shape, though the degrees are suppressed, as compared to the in-umbra spectrum at the same time (Figure \ref{fig:PDspec}).
The time-series sky polarization degrees in the V band and O$_2$ A band exhibit gradual increases after the ingress time toward totality (Figure \ref{fig:PDtimesky}), which are similar to those of the Moon (Figure \ref{fig:PDtime}).
Based only on the snapshot polarization spectra, it is difficult to determine whether the sky polarization was attributed to the polarization of the eclipsed Moon or caused by scattering within the local atmosphere.   
However, the time variation of the sky polarization degrees is unlikely to be explained if the Moon was kept unpolarized because the relative positions of the slits to the Moon were fixed during our observations.  
It seems natural to interpret the sky polarization as  originating  from the polarization of the Moon, which varied with time as shown in Figure \ref{fig:PDtime}.
This interpretation is consistent with the position angles of the sky polarization (Figure \ref{fig:PAtimesky}), which approximately agree with those of the Moon  (Figure \ref{fig:PAtime}).

Based on our error estimates and the interpretation of the observed sky polarization, we judge that our results are legitimate.

\section{Detailed discussion on the first transmission}\label{apx:detail}

\subsection{Analogy with Jupiter's Limb Polarization}\label{apx:Jupiter}
We attempt to explain the observed polarization by multiple scattering, namely, double scattering, during the first transmission.
This idea is noted in association with the polarization pattern on Jupiter \citep{coyn1970}. 
Since the pioneering observations by \cite{lyot1929}, Jupiter has been known to exhibit polarization enhancement on its limb, even though the phase angle is always small ($<11^\circ$) when observed from the Earth.
The degree of the limb polarization is up to 10 \% in the visible wavelengths, whereas that at the center of the disk is less than 1 \% \citep{gehr1969,schm2011}.
The position angles usually have a radial pattern, {which} is vertical to the limb.  
The emergence of the limb polarization was explained by an imbalance between (i) the contribution of light that first scattered to the vertical direction and then scattered to the observer, and (ii) the contribution of light that first scattered to the horizontal direction and then scattered again to the observer \citep{gehr1969,schm2011}.
Grazing incidence to the atmosphere on the limb produces an excess of (ii) over (i), which results in vertical polarization. 
Limb polarization was observed not only for Jupiter but also for Saturn, Uranus, and Neptune \citep{schm2011,schm2006}.

During a lunar eclipse, the Earth is viewed from the Moon as a ring of bright limb.
There are considerable differences between an observation of the Earth's limb and that of Jupiter's limb (and limbs of the other outer planets): 
for instance, the Earth's limb was viewed as transmitted light, whereas Jupiter's limb is observed as reflected light. 
Nonetheless, the polarization of Jupiter's limb may be analogous to the polarization of the Earth's limb, if any,  because the both have a common geometry of grazing incidence to the atmosphere.

Therefore, we examine whether the Earth's limb can be polarized by double scattering, in analogy with Jupiter's limb polarization.
For this purpose, the light after the first Earth transmission is divided into two components, which we label straightforward and  doubly scattered components.
As described below through Appendix {\ref{apx:detail}}, we expect that the former is almost unpolarized and the latter is significantly polarized.

\subsection{Straightforward Component}

The straightforward component refers to the beams that propagate nearly straightforwardly through the atmosphere.
This component contains the direct sunlight and singly scattered sunlight with a scattering angle less than $\sim$2$^{\circ}$ (i.e., nearly forward-scattered light).

The straightforward beams (of whichever type) undergo extinction along the transmission through the atmosphere.
The extinction has been considered to be isotropic with respect to the orientation of electromagnetic oscillations, hence producing no polarization.
Although recent precise ground-based polarimetry discovered that the partially aligned dust in the telluric atmosphere polarizes the incoming stellar light, its degree was 0.005 \% at the maximum \citep{bail2008}.
 
Polarization by refraction of the direct sunlight is roughly evaluated for a uniform plane-parallel atmosphere with a refractive index of the standard air \citep{cidd1996}.
The Fresnel equations give 0.03 \% as the polarization degree of the refracted light in the visible wavelengths.
The polarization by near-forward scattering is also very weak, namely, 0.06 \% for a scattering angle of $2^{\circ}$.

Considering the discussions above, we regard directly transmitted light as unpolarized.

\subsection{Doubly Scattered Component}\label{apx:double}

\begin{figure}[htbp]
\begin{center}
\includegraphics[width=0.67\linewidth]{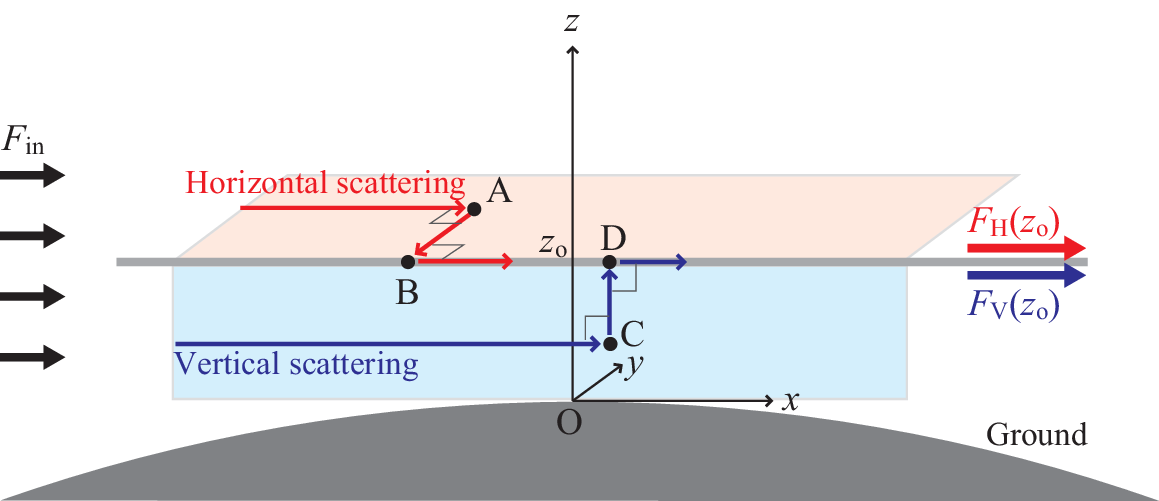}
\caption{Geometry of our double-scattering model.
$F_\mathrm{in}$ is the uniform solar flux incoming to the atmosphere.
Points A and B are the first and second scattering points, respectively, in a horizontal plane (shown in pale red).
Points C and D are the first and second scattering points, respectively, in a vertical plane (shown in pale blue).
$F_\mathrm{H}(z_\mathrm{o})$ is the total horizontally scattered flux outgoing along the horizontal gray line with {a} tangential altitude of $z_\mathrm{o}$.
$F_\mathrm{V}(z_\mathrm{o})$ is the same as $F_\mathrm{H}(z_\mathrm{o})$ except for the vertically scattered flux.
Note that a secondly scattered beam from point B or D may not be exactly in the gray line. 
Some tilt (up to $\sim$2$^\circ$) is necessary in order that the outgoing beam reaches the Moon.
Atmospheric refraction is not illustrated.
}
\label{fig:AtmosScatter}
\end{center}
\end{figure}

In this section, using a simple model, we will demonstrate that the doubly scattered component can be considerably polarized.
As a preparation for our model, we introduce the vertical and horizontal optical thickness together with their ratio, following the formulation by {\cite{fort2005}}.
The vertical optical thickness above an altitude $z$ is written as
\begin{eqnarray}
\tau_\mathrm{V} (z)  =  \int_{z}^{\infty}{\sigma n_z(z)\mathrm{d}z} = \sigma n_z(z)H,
\label{eq:tauV}
\end{eqnarray}
where $\sigma$ is the molecular scattering cross section and $n_z(z)$ is the volume number density of molecules along a vertical line (the $z$-axis in Figure \ref{fig:AtmosScatter}), which is given by $n_z(z)=n_z(0) \mathrm{exp}(-z/H)$. 
$H$ is the scale height derived from 
\begin{eqnarray}
H= \frac{kT}{mg} ,
\label{eq:H}
\end{eqnarray}
where $k$ is Boltzmann's constant, $g$ is the gravitational acceleration, $m$ is the mean mass of an atmospheric molecule, and $T$ is the temperature. 
$H$ is approximately 7--8 km for the Earth.
The horizontal optical thickness along a horizontal line with a tangential altitude of $z$ (parallel to the $x$-axis in Figure \ref{fig:AtmosScatter}) is expressed as
\begin{eqnarray}
\tau_\mathrm{H} (z) =  \int_{-\infty}^{\infty}{\sigma n_x(x,z)\mathrm{d}x} = \sigma n_z(z) \sqrt{2 \pi R H},
\label{eq:tauH}
\end{eqnarray}
where $n_x(x,z)$ is the density of molecules along the horizontal line, which is given by $n_x (x,z) =  n_z(z)\, \mathrm{exp}\left( {-x^2}/2 R H \right)$, where $R$ is the Earth's radius.
From Eqs.\ (\ref{eq:tauV}) and (\ref{eq:tauH}),  we can derive the ratio of the horizontal to the vertical  optical thickness  by
\begin{eqnarray}
\eta \equiv \frac{\tau_\mathrm{H} (z)}{\tau_\mathrm{V} (z)} = \sqrt{\frac{2 \pi R}{H}}.
\end{eqnarray}
Here $\eta$ is approximately 70--80, which means that $\tau_\mathrm{H}$ is much greater than $\tau_\mathrm{V}$. 
$\tau_\mathrm{V}(0)$ is roughly 0.1 for clear atmosphere at  $\lambda = 550$ nm \citep{coff1979}, and thus the vertical optical thickness is always thin. 
In contrast, horizontal optical thickness can be either thin or thick depending on the altitude.

Hereafter, we build a simple model to evaluate the doubly scattered fluxes. 
In order to derive the physical essence, we concentrate on horizontal and vertical double scattering with a scattering angle of $\sim$90$^\circ$, as illustrated in Figure \ref{fig:AtmosScatter}.
The scattering medium is assumed to be atmospheric molecules, and thus Rayleigh scattering is considered.
The effects of clouds are not included in the model but will be briefly discussed.

First, the scattered fluxes are modeled for the situation in which $\tau_\mathrm{H}(z_\mathrm{o})$ is thin ($<1$), where $z_\mathrm{o}$ is the tangential altitude from which the beam of interest exits  (Figure \ref{fig:AtmosScatter}).
With $Z_1$ defined by $\tau_\mathrm{H}(Z_1)  = 1$, we can state $z_\mathrm{o} > Z_1$.
For the purpose of keeping our model as simple as possible, we assume that the optically thin and thick natures are clearly divided at the thickness of unity, though we recognize that in reality the transition is continuous.
The adequacy of this treatment will be checked later.

Let us consider a horizontally scattered beam, which is firstly scattered at point A with a position $(x_\mathrm{A}, y_\mathrm{A}, z_\mathrm{o})$ and secondly at point B with a position $ (x_\mathrm{A}, 0, z_\mathrm{o})$ (see Figure \ref{fig:AtmosScatter}). 
Both points are in a horizontal plane with an altitude $z_\mathrm{o}$.
Note that a ``point'' here refers to a unit volume element of the atmosphere.
The flux of the beam after the second scattering is given by $F_\mathrm{in} \, c \sigma n_\mathrm{A} \, c' \sigma n_\mathrm{B} $, where $F_\mathrm{in}$ is the incident sunlight flux (in units of Wm$^{-2}$$\mu \mathrm{m}^{-1}$, where $\mu \mathrm{m}^{-1}$ is the unit of wavelength intervals),
$c$ and $c'$ are the fraction of perpendicularly scattered energy (in terms of per unit wavelength intervals) in the whole energy scattered by a molecule ($c$ is for unpolarized incident light, and $c'$ is for the completely polarized incident light),
$n_\mathrm{A}$ is the molecular density at point A, and $n_\mathrm{B}$ is that at point B.
We have $n_\mathrm{A}=n_x (\sqrt{x_\mathrm{A}^2+y_\mathrm{A}^2},z_\mathrm{o})$ and $n_\mathrm{B}=n_x (x_\mathrm{A},z_\mathrm{o})$.
Because the total horizontally scattered flux emerging from an altitude $z_\mathrm{o}$ (for $z_\mathrm{o} > Z_1$) is the integrated flux of the beams via point A at any point in the horizontal plane, it is written as
\begin{eqnarray}
F_\mathrm{H}^\mathrm{thin} (z_\mathrm{o}) 
& = & \int_{-\infty}^{\infty} \int_{-\infty}^{\infty} 
F_\mathrm{in}\, c \sigma n_\mathrm{A}\, c' \sigma n_\mathrm{B} \,
\mathrm{d}x_\mathrm{A} \mathrm{d}y_\mathrm{A} \nonumber \\
& = & \frac{\sqrt{2}}{2} F_\mathrm{in} c c' \tau_\mathrm{H}(z_\mathrm{o})^2.
\label{eq:FHthin}
\end{eqnarray}
Eq.~(\ref{eq:tauH}) is used for the transformation.
It is reasonable that the flux of horizontal double scattering at a given altitude, $F_\mathrm{H}^\mathrm{thin} (z_\mathrm{o})$, is proportional to the square of the horizontal thickness at the altitude, $\tau_\mathrm{H}(z_\mathrm{o})^2$.  
 
Evaluation of the vertically scattered flux for $z_\mathrm{o} > Z_1$ can be done in a similar manner to the horizontally scattered flux, but requires separate treatments depending on the altitude of the first scattering point C with a position $(x_\mathrm{C}, 0, z_\mathrm{C})$. 
When point C is within the horizontally thin layer, or for $z_\mathrm{C} > Z_1$, the vertically scattered  flux is
\begin{eqnarray}
F_\mathrm{V}^\mathrm{thin1} (z_\mathrm{o}) 
& = & \int_{Z_1}^{\infty} \int_{-\infty}^{\infty} 
F_\mathrm{in}\, c \sigma n_\mathrm{C} \, c' \sigma n_\mathrm{D} \,
\mathrm{d}x_\mathrm{C} \mathrm{d}z_\mathrm{C} \nonumber \\
& = & \frac{\sqrt{2}}{2} F_\mathrm{in} c c' \tau_\mathrm{H}(z_\mathrm{o}) \tau_\mathrm{V}(Z_1),
\label{eq:FV1}
\end{eqnarray}
where $n_\mathrm{C}$ and $n_\mathrm{D}$ are the molecular densities at the first and second scattering points, C and D, respectively. 
It is natural that the vertically scattered flux from the horizontally thin layer ($z_\mathrm{C} > Z_1$) is proportional to the product of the horizontal thickness at a given altitude, $\tau_\mathrm{H}(z_\mathrm{o})$, and the vertical thickness of the layer, $\tau_\mathrm{V}(Z_1)$.

When point C is within the horizontally thick layer, or for $z_\mathrm{C} \le Z_1$, the incident sunlight can no longer penetrate through the atmosphere.
With  $X_\mathrm{u}$ as the $x$ coordinate where the horizontal optical depth reaches unity (thus $\int_{-\infty}^{X_\mathrm{u}} \sigma n_x\mathrm{d}x = 1$), we have
\begin{eqnarray}
F_\mathrm{V}^\mathrm{thin2} (z_\mathrm{o}) 
& = & \int_{0}^{Z_1} \int_{-\infty}^{X_\mathrm{u}} 
F_\mathrm{in}\, c \sigma n_\mathrm{C} \, c' \sigma n_\mathrm{D} \,
\mathrm{d}x_\mathrm{C} \mathrm{d}z_\mathrm{C} \nonumber \\
& \cong & (- \log u + u - 1) F_\mathrm{in} c c' \tau_\mathrm{H}(z_\mathrm{o}) \tau_\mathrm{V}(Z_1),
\label{eq:FV2}
\end{eqnarray}
where $u \equiv \left(2-\sqrt{2} \right)/4$ and thus $- \log u + u - 1 \cong 1.1$. 
Here, for the reason of keeping our calculation analytical, we have replaced a Gaussian function $\exp\left( -a x^2 \right) $ by a rectangular function with a width of $2\sqrt{\pi/a}$ and a height of $1/2$.
This corresponds to a physical treatment in which we flatten the Gaussian distribution of $n_x$ with a constant value within a limited $x$ range, whilst the total number of molecules is conserved.

By comparing Eq.~(\ref{eq:FV1}) with Eq.~(\ref{eq:FV2}), we find $F_\mathrm{V}^\mathrm{thin1}: F_\mathrm{V}^\mathrm{thin2} \cong 1:1.5$.
Despite much greater molecular densities below $Z_1$ than those above $Z_1$, $F_\mathrm{V}^\mathrm{thin2}$ is comparable to $F_\mathrm{V}^\mathrm{thin1}$.
This can be interpreted as follows:
with the increasing $\tau_\mathrm{H}$ (decreasing $z_\mathrm{C}$), $X_\mathrm{u}$ moves toward the horizontal end (to the negative $x$ direction or the left side in Figure \ref{fig:AtmosScatter}); 
this means that the effective $x$ range for the scattering point D becomes more limited, and thus the secondly scattered flux decreases with decreasing $z_\mathrm{C}$;
consequently, $F_\mathrm{V}^\mathrm{thin2}$ is not too great compared to $F_\mathrm{V}^\mathrm{thin1}$.

Because the net flux of vertically scattered light is the sum of Eqs.~(\ref{eq:FV1}) and (\ref{eq:FV2}), it is written as
\begin{eqnarray}
F_\mathrm{V}^\mathrm{thin} (z_\mathrm{o}) 
\cong  (- \log u -u ) F_\mathrm{in} c c' \tau_\mathrm{H}(z_\mathrm{o}) \tau_\mathrm{V}(Z_1),
\label{eq:FVthin}
\end{eqnarray}
where $ - \log u -u$  is approximately 1.8. 

Next, we model the scattered fluxes when $\tau_\mathrm{H}(z_\mathrm{o})$ is thick ($\ge1$), or for $z_\mathrm{o} \le Z_1$.
In this condition, light extinction during the propagation cannot be ignored.
This effect diminishes the emergent scattered fluxes (for both horizontal and vertical scattering) by a factor of approximately $\exp \left (- \tau_\mathrm{H}(z_\mathrm{o}) \right)$.
With decreasing $z_\mathrm{o}$ from $Z_1$, the extinction becomes double-exponentially more effective. 
On the other hand, the scattered energy (or number of photons) migrated to the horizontal line with a tangential altitude of $z_\mathrm{o}$ is almost saturated for $z_\mathrm{o} < Z_1$, and thus it does not vary much with decreasing $z_\mathrm{o}$.
Therefore, the emergent scattered fluxes ($F_\mathrm{H}^\mathrm{thick} (z_\mathrm{o}), F_\mathrm{V}^\mathrm{thick} (z_\mathrm{o})$ for horizontal and vertical scattering, respectively) can be modeled  as
\begin{eqnarray}
F_\mathrm{H}^\mathrm{thick} (z_\mathrm{o}) 
& \cong  & F_\mathrm{H}^\mathrm{thin}(Z_1) \exp \left (- \tau_\mathrm{H}(z_\mathrm{o}) \right),
\label{eq:FHthick} \\
F_\mathrm{V}^\mathrm{thick} (z_\mathrm{o}) 
& \cong  & F_\mathrm{V}^\mathrm{thin}(Z_1) \exp \left (- \tau_\mathrm{H}(z_\mathrm{o}) \right).
\label{eq:FVthick}
\end{eqnarray}

\begin{figure}[htbp]
\begin{center}
\includegraphics[width=0.7\linewidth]{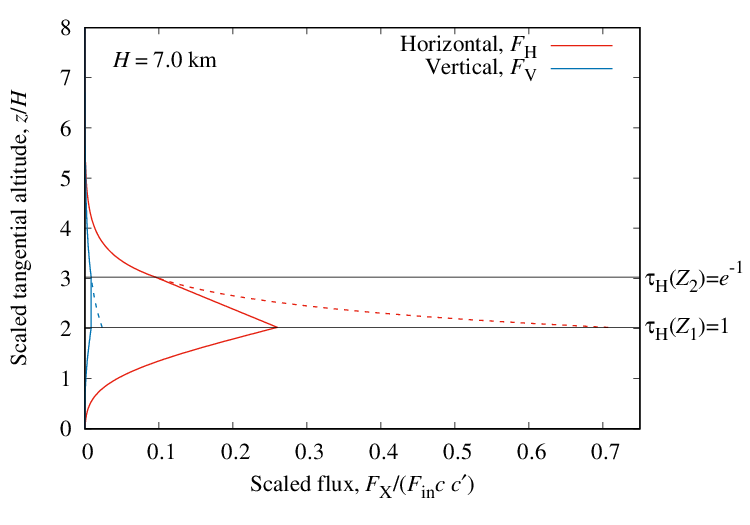}
\caption{Altitudinal profile of doubly scattered fluxes.
The vertical axis shows the tangential altitude, scaled by the scale height $H$.
The horizontal axis shows the horizontally or vertically scattered fluxes, scaled by $F_\mathrm{in}cc'$. 
The solid lines are our model fluxes ($F_\mathrm{X}$; $\mathrm{X}$ represents $\mathrm{H}$ or $\mathrm{V}$), in which extinction in a layer with intermediate horizontal optical thickness is taken into consideration.
$F_\mathrm{H}$ is drawn in red and $F_\mathrm{V}$ is in blue.
The dashed lines are the extensions of model fluxes for the optically thin layer ($F_\mathrm{X}^\mathrm{thin}$).
These model fluxes are calculated with settings $\tau_\mathrm{V}(0)=0.1$  \citep[the value at $\lambda = 550$ nm,][]{coff1979} and $H=7.0$ km.
The horizontal lines show the altitudes $Z_1$ and $Z_2$, where $\tau_\mathrm{H}$ equals 1 and $e^{-1}$, respectively. 
}
\label{fig:polflux}
\end{center}
\end{figure}

$F_\mathrm{X}^\mathrm{thin}(z)$ and $F_\mathrm{X}^\mathrm{thick}(z)$ (X is H or V) are plotted in Figure \ref{fig:polflux} (hereafter $z_\mathrm{o}$ is replaced by $z$). 
At $z=Z_1$, there is a disconnection between $F_\mathrm{X}^\mathrm{thin}(z)$ and $F_\mathrm{X}^\mathrm{thick}(z)$.
This is because we treated $Z_1$ as a clear divide between thin and thick layers and neglected light extinction for $z>Z_1$.
In reality, however, extinction is effective for a $z$ range where $\tau_\mathrm{H}(z)$ is intermediate, say, a few tenths to unity.
In order to reduce the flux overestimate for optically intermediate altitudes, we apply a linear connection between $Z_1$ and $Z_2$ (defined by $\tau_\mathrm{H}(Z_2) = 1/e$) as shown in Figure \ref{fig:polflux}.

In summary, we have horizontally and vertically scattered fluxes:
\begin{eqnarray}
F_\mathrm{H}(z) & = & \left\{
\begin{array}{lcll}
F_\mathrm{H}^\mathrm{thin}(z) & = & \frac{\sqrt{2}}{2} F_\mathrm{in} cc' \tau_\mathrm{H}(z)^2 & (z > Z_2)\\
F_\mathrm{H}^\mathrm{thick}(z) & \cong & F_\mathrm{H}^\mathrm{thin}(Z_1) \exp \left (- \tau_\mathrm{H}(z) \right) & (z \le Z_1)
\end{array}
\right. ,
\label{eq:FH}
\end{eqnarray}
and
\begin{eqnarray}
\lefteqn{F_\mathrm{V}(z)} \nonumber \\
 & = & \left\{
\begin{array}{lcll}
F_\mathrm{V}^\mathrm{thin}(z) & \cong & (- \log u -u ) F_\mathrm{in} c c' \tau_\mathrm{H}(z) \tau_\mathrm{V}(Z_1) & (z > Z_2)\\
F_\mathrm{V}^\mathrm{thick}(z) & \cong & F_\mathrm{V}^\mathrm{thin}(Z_1) \exp \left (- \tau_\mathrm{H}(z) \right) & (z \le Z_1)
\end{array}
\right. ,
\label{eq:FV}
\end{eqnarray}
respectively.
For a range $Z_1 < z \le Z_2$, a linear connection is applied. 

By comparing  Eqs.~(\ref{eq:FH}) and (\ref{eq:FV}), we can derive the ratio of the horizontally scattered flux to the vertically scattered flux as
\begin{eqnarray}
\frac{F_\mathrm{H}(z)}{F_\mathrm{V}(z)} & \cong & \left\{
\begin{array}{lcll}
\frac{\tau_\mathrm{H}(z)}{2.5 \tau_\mathrm{V}(Z_1)} & \cong & 30\; \tau_\mathrm{H}(z) & (z > Z_1)\\
\frac{\tau_\mathrm{H}(Z_1)}{2.5 \tau_\mathrm{V}(Z_1)} & \cong & 30 & (z \le Z_1)
\end{array}
\right. .
\label{eq:Fratio}
\end{eqnarray}
For $z < 5.5 H \cong 40$ km (for $\lambda =$ 550 nm), $F_\mathrm{H}$ is larger than $F_\mathrm{V}$.
At $z = Z_1$ where both  $F_\mathrm{H}$ and $F_\mathrm{V}$ are at the maximum (see Figure \ref{fig:polflux}), $F_\mathrm{H}$ is approximately 30 times larger than $F_\mathrm{V}$.
Note that the flux contribution from altitudes $z \ge 5.5 H$ is negligible, as is obvious from Figure \ref{fig:polflux}. 

We calculate the vertically integrated emergent fluxes (in units of W m$^{-1}$$\mu \mathrm{m}^{-1}$),
\begin{eqnarray}
\Phi_\mathrm{H} & \equiv  & \int_{0}^{\infty} F_\mathrm{H}(z) \mathrm{d}z
\cong \frac{\sqrt{2}}{4}\left( \frac{3}{e} + \frac{2}{e^2} \right) F_\mathrm{in} c c' H,
\label{eq:PhiH}\\
\Phi_\mathrm{V} & \equiv  & \int_{0}^{\infty} F_\mathrm{V}(z) \mathrm{d}z
\cong \frac{3  (- \log u -u )}{e \eta} F_\mathrm{in} c c' H,
\label{eq:PhiV}
\end{eqnarray}
where $\int_{0}^{Z_1} F_\mathrm{X}^\mathrm{thick}(z) \mathrm{d}z$ is approximated by a triangular area with a base of $F_\mathrm{X}^\mathrm{thick}(Z_1)$ and a height of $2H$.
We have $\Phi_\mathrm{H}/\Phi_\mathrm{V} \cong 0.25\eta$, which is approximately 20.
Hence, $\Phi_\mathrm{H}$ is much greater than $\Phi_\mathrm{V}$.
The polarization degree of the doubly scattered component is estimated to be $\sim$90 \% in our model.
Because horizontally scattered flux is dominant, the emergent scattered flux is polarized in the vertical direction.
Therefore, the ring of the Earth's limb as viewed from the Moon is expected to have a radial pattern of polarization. 

The radial pattern of polarization is in common with Jupiter's limb, as described in section \ref{apx:Jupiter}.
Our estimate also seems consistent with the polarization pattern of the sunlit sky viewed from the Earth ground \citep{berr2004}: 
the observational and theoretical results had vertical polarization of the sky near the Sun on the horizon; the polarization of the sky near the Sun was explained by the effect of double scattering.

Although each element in the Earth's limb is expected to be polarized, the sum of the whole limb would be unpolarized if the all the limb elements had an equal polarized flux in terms of the absolute value (i.e., the vectors of the polarized flux had a perfectly point-symmetric pattern).
Our detection of polarization on the eclipsed Moon requires some sort of inhomogeneity along the limb, because the observed light is a reflection of the integrated Earth's limb.
As our observed position angle of polarization is roughly in the east--west orientation (Figure \ref{fig:PAtime}), we expect an excess of polarized flux at the equatorial regions compared to that at the polar regions.  

We have not yet determined for certain what causes the inhomogeneity of polarized flux.
Although a fully quantitative discussion is the beyond the scope of this work, there are a few  qualitatively possible explanations.
From Eqs. (\ref{eq:PhiH}) and (\ref{eq:PhiV}), we obtain the polarized flux of scattered light from the atmosphere above a point on the limb: $P_\mathrm{db}\Phi_\mathrm{db} \equiv \Phi_\mathrm{H} - \Phi_\mathrm{V} \cong \Phi_light\mathrm{H}$, where $P_\mathrm{db}$ and $\Phi_\mathrm{db}$ are the polarization degree and the flux of the doubly scattered component, respectively.
$\Phi_\mathrm{H}$ is proportional to $H$ (Eq.~(\ref{eq:PhiH})), which is proportional to the temperature $T$ (Eq.~(\ref{eq:H})).
Hence, $P_\mathrm{db}\Phi_\mathrm{db}$ is estimated to be proportional to $T$.
According to the April data in the COSPAR International Reference Atmosphere (CIRA-86), pressure-weighted vertically averaged temperatures at the equator (0$^\circ$ latitude) and the poles (80$^\circ$N and 80$^\circ$S latitudes) are $\sim$265 K and $\sim$240 K, respectively. 
Therefore, we can expect an excess of $\sim$10 \% of $P_\mathrm{db}\Phi_\mathrm{db}$ at the equatorial regions compared to that at the polar regions.
The inhomogeneous polarized flux due to the latitudinal variation of temperature may account for part of the observed polarization.

Alternatively or additionally, local coverage of clouds, or any diffuse media, may also explain the inhomogeneous polarized flux along the limb.
The photographs of Earth's limb taken by Surveyor III showed that the local clouds on the beam path strongly reduce the local brightness of the limb \citep{shoe1968}.
Hence, if the cloud coverage near the poles was greater than that near the equator at the time of our observations, an excess of polarized flux at the equator may have been produced.
As shown in Figure \ref{fig:polflux}, the most contributing altitude to the polarized flux is $z=Z_1\cong 2H \sim 15$ km, which is in the lower stratosphere.
Although the stratosphere is rarely cloudy owing to its dryness, the stratosphere above the polar regions is known to occasionally contain clouds, which are called polar stratospheric clouds or PCSs.
We checked a global cloud map on our observing date (not exactly at the time of our observations), obtained by the MODIS instrument aboard the Terra satellite.\footnote{\url{https://worldview.earthdata.nasa.gov}}
Visual inspection of the map suggested more clouds on the polar regions than the equator.
This may support our hypothesis; nonetheless, more comprehensive investigation is necessary  to prove it.

The latitudinal inhomogeneity may be either temporary or constant.
Further monitoring of lunar eclipse polarization will help to identify the cause of the inhomogeneity.

\subsection{Consistency with the Observations}\label{apx:consist}
Overall, the observed in-umbra polarization degrees decrease with increasing wavelength (Figure \ref{fig:PDspec}).
This wavelength dependence is consistent with our explanation of polarization caused by double scattering during the first transmission.
Recall that the transmitted light is the sum of the unpolarized straightforward and polarized doubly scattered components.
Hence, we have
\begin{eqnarray}
P_\mathrm{e} =  \frac{P_\mathrm{db}\Phi_\mathrm{db}}{\Phi_\mathrm{db} + \Phi_\mathrm{fwd}}
\cong \frac{\Phi_\mathrm{H}}{\Phi_\mathrm{H} + \Phi_\mathrm{fwd}},
\label{eq:Pe}
\end{eqnarray}
where $P_\mathrm{e}$ is the polarization degree of an element on the Earth's limb as observed from the Moon and $\Phi_\mathrm{fwd}$ is the vertically integrated flux of the straightforward component.
We have $P_\mathrm{db}\Phi_\mathrm{db}  \cong \Phi_\mathrm{db}  \cong \Phi_\mathrm{H}$.
Eq.~(\ref{eq:PhiH}) tells us that $\Phi_\mathrm{H}$ has the same wavelength dependence as the $F_\mathrm{in}$.
As we have $\Phi_\mathrm{fwd} \propto F_\mathrm{in} {\cal T} $, where ${\cal T}$ is the atmospheric transmittance, the wavelength dependence of the observed polarization can be expressed in the form $P_\mathrm{e}(\lambda) = \left(1 + a {\cal T}(\lambda) \right)^{-1}$, where $a$ is a wavelength-independent constant.
Note that $F_\mathrm{in}$ is canceled out.
As is well known as the reason for the red color of the eclipsed Moon, ${\cal T}$ increases rapidly with increasing wavelength.
Therefore, the polarization degree of the in-umbra Moon rapidly decreases with increasing wavelength (Figure \ref{fig:PDspec}).

The existence of enhanced features on the in-umbra polarization degree spectrum (Figure \ref{fig:PDspec}) is explained in the same manner as the overall wavelength dependence.
At molecular absorption wavelengths, a ${\cal T}$ spectrum has a decreased feature.
Therefore, as the inverse of the ${\cal T}$ spectrum, the polarization degree spectrum has an enhanced feature.
The distinct contrast of the feature at the $\sim$760 nm O$_2$ absorption wavelengths is interpreted as follows.
At the continuum wavelengths, unpolarized $\Phi_\mathrm{fwd}$ is dominant as compared to polarized $\Phi_\mathrm{db}$, owing to a great atmospheric transmittance at the red wavelengths.
In contrast, at the absorption wavelengths, the contribution of $\Phi_\mathrm{db}$ should be significant owing to the great depression of $\Phi_\mathrm{fwd}$.
Hence, the polarization degree at the continuum is very low, whereas that at the absorption band can be distinctively high.

The observed polarization degrees at the V band and the O$_2$ A band increased from the time of target ingress toward the time of the greatest eclipse (Figure \ref{fig:PDtime}).
This time variation is also consistent with our explanation.
As the Moon moves deeper in the umbra, the Moon becomes darker.
The V-band surface brightness of the eclipsed Moon darkens by a factor of tens with respect to the brightness at the ingress time \citep{seki1980}.
The decrease {in} the brightness should be attributed to that of the straightforward component $\Phi_\mathrm{fwd}$, rather than the doubly scattered component $\Phi_\mathrm{db}$.
As the Moon approaches the umbra center, $\Phi_\mathrm{fwd}$ decreases because the direct sunlight subcomponent traces a lower and thicker atmosphere, which is a result of the requirement of a greater refraction angle to be directed to the Moon.
In contrast, $\Phi_\mathrm{db}$ should remain virtually constant during the movement of the Moon 
because the flux of Rayleigh scattering varies as little as 0.1 \% with a direction change of  a range of $\sim$2$^\circ$ \citep{hapk2012}.
Therefore, the observed increasing polarization degrees after the ingress time to the mid-time can be explained by the decreasing $\Phi_\mathrm{fwd}$ and the nearly constant  $\Phi_\mathrm{db}$  ($\cong \Phi_\mathrm{H}$) (see Eq.~(\ref{eq:Pe})).

\subsubsection{Limitations}
We should mention some limitations of our simple model for double scattering.
Because our model contains simplifications and approximations, it is not sufficient to be used for an estimate of absolute fluxes, namely, $\Phi_\mathrm{H}$, $\Phi_\mathrm{V},$ and $\Phi_\mathrm{db}$. 
One of the concerns arising from this may be whether the polarized flux of double scattering can be significant enough within the total transmitted flux, which is considered here as the sum of direct, singly scattered, and doubly scattered components. 

Although the direct sunlight transmitted through the Earth's atmosphere has been believed to be the dominant component in the brightness of the eclipsed Moon, recent studies  showed that the singly scattered component can be comparable to, or even surpass, the direct component, especially at wavelengths shorter than 600 nm  \citep{garc2011a,garc2011b}.
They did not consider any higher-order scattering, but an evidence suggesting the significance of double scattering is provided from facts on polarized sunlit sky observed from the ground.
{The polarization} pattern over the sunlit sky cannot be explained by only single scattering, but the contribution of multiple scattering, namely, double scattering, is also required \citep{chan1954,coff1979,berr2004}. 
The theoretical calculation by \cite{chan1954} showed that a sky region near the Sun on the horizon has a polarization degree of $\sim$20 \%, which  should be attributed to double scattering, because near-forward single scattering cannot produce such strong polarization. 
If we observe the sunlit sky from the Moon, the sky polarization may be larger owing to a larger optical thickness for the transmission than that for ground-based observations, which means singly and doubly scattered fluxes may be comparable.
Combining previous knowledge about the brightness of the eclipsed Moon (possibly, direct sunlight $\sim$ single scattering) and polarization of sunlit sky (possibly, single scattering $\sim$ double scattering), we expect that fluxes of direct sunlight, single scattering, and double scattering can be comparable to each other and that their sum may have a polarization percentage of several tens, especially  at wavelengths shorter than 600 nm.

Another limitation in our model is that we consider only Rayleigh scattering by gas molecules as the polarizing source.
This assumption is reasonable because the polarization of the sunlit sky, as observed from the ground, is excellently reproduced by the theory based on Rayleigh scattering \citep{chan1954}.
Nonetheless, polarization by aerosol scattering may also explain the observed polarization.
$\Phi_\mathrm{db}$ ($\cong \Phi_\mathrm{H}$) in our model (Eq.~(\ref{eq:PhiH})) is missing a term representing the wavelength dependence of Rayleigh scattering (namely, $\sigma$).\footnote{This is because a different $\sigma$ makes a different altitude of the thin/thick boundary ($Z_1$) and thus vertically shifts the profile of $F_\mathrm{H}(z)$ in Figure \ref{fig:polflux}, but does not alter the vertical integral.} 
The wavelength dependence of the in-umbra polarization degree is basically determined by that of the transmittance for unpolarized light $\Phi_\mathrm{fwd}$.
Therefore, our model does not exclude polarization by non-Rayleigh scattering as the cause of polarization.
In the case where aerosol polarization is effective, the distribution of aerosols may explain the inhomogeneity of the polarized flux along the Earth's limb.

In spite of these limitations, our model for double scattering is useful to demonstrate the overwhelming excess of horizontally scattered flux over vertically scattered flux and to identify the determinant factors of the polarized flux.

\section{Expected Fractional Polarization during a Transit of an Earth-like Exoplanet}\label{apx:estimate}

Based on our observed results, we make a rough estimate of fractional polarization during a transit of an Earth-like exoplanet.
We detected 2-3 \% polarization of the in-umbra Moon.
Considering a depolarization factor of $\sim$1/3 \citep{doll1957,bazz2013}, polarization of the net (ring-integrated) transmitted light through the Earth's atmosphere should have been $\sim$6-9\% if we had observed the Earth from the Moon.
We assume a 10\% polarization of the Earth's transmitted light ($P_\mathrm{earth}$) for the following discussion.

According to a lunar eclipse observation by \cite{seki1980}, the V-band surface brightness was darker by  $\sim$10 mag when the Moon was in the umbra than when it was outside of the penumbra.   
This means $F_\mathrm{earth}/F_\mathrm{sun} \sim 10^{-4}$, where $F_\mathrm{earth}$ is the net flux of Earth's transmitted light as observed from the Moon and $F_\mathrm{sun}$ is direct sunlight (without transmission through the Earth's atmosphere).
Hence, the ratio of the polarized flux to the sunlight is $P_\mathrm{earth}F_\mathrm{earth}/F_\mathrm{sun} \sim10^{-5}$.

We need to be careful of a difference between the geometry of an Earth's transit as observed from the Moon and that of an exoplanet's transit as observed from the Earth. 
Let us imagine what would happen if we pushed the Moon further away from the Earth.
As the original distance of the Moon from the Earth, $l$, becomes a larger distance, $l'$, the polarized flux observed from the Moon will decrease to $P_\mathrm{earth}F_\mathrm{earth} (l'/l)^{-2}$.
In the meantime, the received solar flux will be $F_\mathrm{sun} \left( (L+l')/L \right)^{-2}$, where $L$ is the original distance between the Sun and the Moon.
Hence, the ratio of the polarized flux to the solar flux, as observed from the further point, will be  
\begin{eqnarray}
\frac{P_\mathrm{earth}F_\mathrm{earth}}{F_\mathrm{sun}}  \left( \frac{l'}{l} \frac{L}{L+l'} \right)^{-2}.
\end{eqnarray}
An observation of an exoplanetary transit corresponds to a situation where $l' \rightarrow \infty$; thus, the fractional polarization during the transit, $P_\mathrm{exo}$, can be written as 
\begin{eqnarray}
P_\mathrm{exo} = \frac{P_\mathrm{earth}F_\mathrm{earth}}{F_\mathrm{sun}}  \left( \frac{L}{l} \right)^{-2}.
\end{eqnarray}
Putting actual values ($L\cong1.5 \times 10^8$ km, $l \cong 4 \times 10^5$ km) into this equation gives $P_\mathrm{exo} \sim 7 \times 10^{-11}$.
We have to say that detecting polarization during the transit of an Earth-like exoplanet is extremely challenging with current and near-future technologies.

\end{document}